\documentclass[traditabstract]{aa} 
\usepackage{graphicx}
\usepackage{color}
\usepackage{natbib}
\bibpunct{(}{)}{;}{a}{}{,} %
\usepackage{txfonts}
\def\aap{A\&A}

\begin{document}

   \title{VLTI/AMBER observations of the binary B[e] supergiant HD 327083\thanks{Based on observations conducted at the European Southern Observatory, Paranal, Chile, which were obtained as part of the program 383.C-0166.}}

   \author{H.E. Wheelwright
          \inst{1,2}
          \and
         W.J. de Wit\inst{3}	
         \and
         R.D. Oudmaijer\inst{2}
         \and
         J.S. Vink\inst{4}
         }

\institute{
Max-Planck-Institut f\"{u}r Radioastronomie, Auf dem H\"{u}gel 69, 53121 Bonn, Germany\\\email{hwheelwright@mpifr-bonn.mpg.de}
    \and
School of Physics and Astronomy, University of Leeds, Leeds LS2 9JT, UK
    \and 
European Southern Observatory, Alonso de Cordova 3107, Vitacura, Santiago, Chile
\and  
Armagh Observatory, College Hill, Armagh BT61 9DG, Northern Ireland
}

   \date{Received July 26, 2011; accepted December 04, 2011}

 
   \abstract {\object{HD 327083} is a luminous B type star which
     exhibits emission lines {{and an infrared excess}} and is therefore
     classified as a supergiant B[e] star. In addition, the star is
     the primary of a close binary system. It is not clear whether the
     B[e] behaviour of HD 327083 is related to its binarity or its
     evolutionary state. Here we address this issue by studying its
     circumstellar environment with high spatial resolution. To this
     end, we have observed HD 327083 with the VLTI and AMBER in the
     medium resolution $K$-band setting. $\mathrm{^{13}CO}$ bandhead
     emission is detected, confirming HD 327083 is a
     {{post-main sequence}} object. The observations spatially resolve the
     source of the NIR continuum and the Br$\gamma$ and CO line
     emission. In addition, differential phase measurements allow us
     to probe the origin of the observed Br$\gamma$ emission with
     sub-mas precision. Using geometrical models, we find that the
     visibilities and closure phases suggest that the close binary
     system is surrounded by a circum-binary disk. We also find that
     in the case of the binary HD 327083, the relative sizes of the
     continuum and Br$\gamma$ emitting regions are different to those
     of a single supergiant B[e] star where
     the standard dual outflow scenario is thought to apply. These
     findings are consistent with the hypothesis that the mass loss of
     HD 327083 is related to its binary nature.}

   \keywords{Techniques: high angular resolution --
             Techniques: interferometric --
	     Stars: early type --
	     Stars: emission-line, Be --
	     Stars: mass-loss--
             Stars: individual: HD 327083}

\titlerunning{The binary sgB[e] HD 327083 resolved with the VLTI/AMBER}
\authorrunning{H.E. Wheelwright et al.}
   \maketitle
%

\section{Introduction}

Massive stars play a pivotal role in many areas of
astrophysics. Consequently, it is important to understand their
evolution and how they interact with their surroundings. In turn,
understanding the evolution of massive stars requires knowledge of how
they lose mass and how this changes with evolutionary state \citep[see
e.g.][]{Puls2008}. Furthermore, constraining the geometry of massive
star outflows is of great importance for massive star evolution models
that include stellar rotation, especially with respect to the removal
of angular momentum. Supergiant B[e] {{(sgB[e])}} stars are key
objects in this regard as they are massive objects in a late
evolutionary stage which also exhibit signs of enhanced mass loss
\citep[see e.g.][]{Lamers1998}. However, the mechanism responsible for
the mass loss of sgB[e] stars is still not fully understood.

\smallskip

The primary characteristics of sgB[e] stars are an optical spectrum
with prominent emission lines and an excess of infrared continuum
emission. Typical emission lines include: Balmer lines; low excitation
lines of metals such as Fe{\sc{ii}}; and forbidden lines of
Fe{\sc{ii}} and O{\sc{i}}. The low excitation emission lines are
generally relatively narrow ($\sim$10~s of
$\mathrm{km\,s^{-1}}$). Typically, sgB[e] stars do not exhibit
absorption features in the optical, but do display absorption lines
due to ions such as Si{\sc{iv}} and C{\sc{iv}} in the UV. These
features are high excitation lines, and are generally much broader
than the low excitation emission lines (e.g. 1000 s of
$\mathrm{km\,s^{-1}}$). As a result, the spectra of sgB[e] have a
hybrid appearance.

\smallskip

To explain these characteristics, \citet{Zickgraf1985} proposed a
model in which sgB[e] stars drive two distinct outflows: a fast wind
in the polar regions and a slower, and denser outflow in the
equatorial region. In this scenario, the narrow line emission of
sgB[e] stars originates in the dense equatorial wind while the broad
absorption lines trace the fast polar wind. This model reproduces the
range of line profiles exhibited by sgB[e] stars but the origin of the
equatorial outflow is still uncertain \citep[see e.g.][]{Puls2008}. It
has been proposed that when stars become cooler than $\sim$25\,000~K,
an increased Fe opacity might lead to bi-stable outflows in the
equatorial direction \citep{Pelupessy2000}. Alternatively, rapid
rotation and wind compression \citep[see
e.g.][]{Bjorkman1993,Cure2005} may also contribute to equatorial
enhancements. These models need to be tested against observational
constraints, of which there are currently few examples.

\smallskip

Regardless of its origin, is generally thought that this dense,
disk-like equatorial outflow contains warm dust which is responsible
for the infrared excess of sgB[e] stars \citep[see
e.g.][]{DS2011}. However, it has been shown that simple models of the
dust emission from equatorial disks struggle to simultaneously
reproduce the spectral energy distribution of the sgB[e] star R126 in
the optical/near infrared (NIR) and the mid infrared \citep{P2003}. As
a result, it has been suggested that free-free and free-bound emission
from the polar wind can also contribute to the infrared excess of
sgB[e] stars \citep{Kraus2007}. Consequently, understanding the
geometry traced by the NIR excess of sgB[e] stars is not trivial.

\smallskip

Optical interferometry is one of the few techniques that can probe the
environments of sgB[e] stars to study their circumstellar geometries
in detail. For example, \citet{deSouza2007} used the VLTI to study the
sgB[e] star CPD$-$57$^{\circ}$~2874. They found that the sizes of the
observed NIR continuum, Br$\gamma$ emission and mid infrared (MIR)
continuum are different. Furthermore, they also found that the
orientation of the near and mid infrared emission regions on the sky
differs. The MIR emission was later shown to trace a dusty disk-like
structure \citep{DS2011}. These findings are broadly consistent with
the dual outflow scenario of \citet{Zickgraf1985}. However, these
observations do not constrain the origin of the equatorial outflow. In
a more recent use of the VLTI, \citet{Millour2009} used the AMBER
instrument to reconstruct images of the sgB[e] candidate HD 87643 with
high spatial resolution. The images reveal a previously undetected
binary companion. \citet{Millour2009} speculate that the complex
circumstellar environment of the star has been shaped by binary
interactions, suggesting binarity may play a role in the sgB[e]
phenomenon.

\smallskip

Despite the progress made with such observations, a general
understanding of sgB[e] stars has yet to emerge. This is principally
because the number of objects studied with the required spatial and
spectral resolution is still very small. To increase this sample, and
thus address the uncertainty regarding the sgB[e] phenomenon, we
present VLTI/AMBER observations of an additional sgB[e] star, HD
327083 (ALS 3992, CPD-40 7757, MWC 873, He 3-1359). This object is a
Galactic sgB[e] candidate \citep[][]{Kraus2009CO} and is known to
exhibit optical emission lines characteristic of sgB[e] stars
\citep[H{\sc{i}}, He{\sc{ii}} and Fe{\sc{ii}}, see
e.g.][]{Carlson1979}. In addition, it also exhibits CO overtone
bandhead emission at $\mathrm{\sim2.3 \mu m}$
\citep{McGregor1988}. This emission requires hot, dense conditions and
thus could originate in the proposed equatorial outflow of sgB[e]
stars. Therefore, this emission is consistent with the object's
classification as a sgB[e] star \citep[see][]{Kraus2009CO}.

\smallskip

\citet{Lopes1992} suggest that HD 327083 has an intrinsic luminosity
of $\mathrm{10^6}$~{{${L_{\odot}}$}}, making it one of the most
luminous objects in the Galaxy. This luminosity was calculated for a
distance of 5~kpc, which \citet{Lopes1992} estimate from the
equivalent width of the Na {\sc{i}} lines in the object's
spectrum. This high value is subject to a factor of 2 error, as
discussed in \citet{Millour2009}. Also, as noted by \citet{Miro2003},
the interstellar Na {\sc{i}} lines of this object are saturated,
resulting in an overestimation of the distance to it. Based on the
radial velocity of interstellar features in the object's spectrum,
these authors suggest it is located in the Sagittarius spiral arm and
thus assign a distance of $d=1.5 \pm 0.5$~kpc. We adopt this
distance. At this distance, the luminosity will decrease by an order
of magnitude, but is still sufficient to satisfy the criterion of
\citet{Lamers1998} for classification as a sgB[e] star.

\smallskip

Currently, the reason for the B[e] behaviour of HD 327083 is still
uncertain. Based on optical spectroscopy and a NLTE model of an
expanding atmosphere, \citet{Machado2003} suggest that the object may
be close to the {{luminous blue variable phase}}. In the only other
in-depth study of this object, \citet{Miro2003} detected an unresolved
binary companion via radial velocity variations. These authors suggest
that the system is close enough to interact. In this case, the
circumstellar material evidenced by the object's infrared excess may
be the result of binary interactions, as proposed for lower luminosity
B[e] stars \citep[see][]{Miro2007}. Consequently, the evolutionary
state of HD 327083 is still a matter of debate and it is not clear
whether it conforms with the standard sgB[e] scenario or not. To
rectify this, we have observed HD 327083 with the VLTI to probe the
source of the NIR line and continuum emission on milli-arcsecond (mas)
scales and constrain its circumstellar geometry and mass loss.

\smallskip

This paper is structured as follows. Section \ref{obs_and_data}
presents the VLTI and AMBER observations, in addition to the data
reduction process used. We then present the results in {{Sect.}}
\ref{res} and discuss their implications in {{Sect.}}
\ref{disc}. Finally, we conclude the paper in {{Sect.}} \ref{conc}.

\section{Observations and data reduction}

\label{obs_and_data}

\begin{center}
  \begin{figure}
    \begin{center}
      \includegraphics[width=0.375\textwidth]{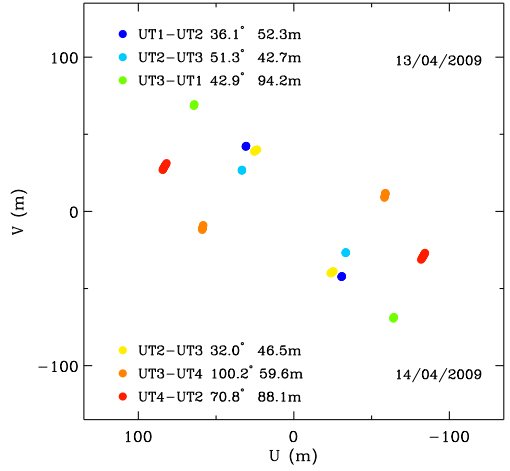}
      \caption{The projected baselines of the observations of HD 327083.\label{uv_cov}}
    \end{center}
  \end{figure}
\end{center}

HD 327083 was observed with the VLTI and AMBER
\citep[see][]{Petrov2007} on the $\mathrm{13^{th}}$ and
$\mathrm{14^{th}}$ of April 2009. During observations on the
$\mathrm{13^{th}}$, light from the unit telescopes UT1, UT2 and UT3
was combined while on the $\mathrm{14^{th}}$ the UT2, UT3 and UT4
telescopes were used. The fringe tracker FINITO was used and the
exposure times were 187~ms on the $\mathrm{13^{th}}$ and 300~ms on the
$\mathrm{14^{th}}$. On the $\mathrm{13^{th}}$, 1000 spectrally
dispersed interferograms were recorded while 1800 such frames were
recorded the following night. The data were obtained using the medium
resolution $K$ setting which has a spectral resolution of R~=~1500 or
$\mathrm{\sim200~km\,s^{-1}}$. The use of FINITO allowed us to record
the entire spectral range and thus we observed both Br$\gamma$
emission at 2.16~$\mathrm{\mu m}$ and CO bandhead emission at
$\sim$2.3~$\mathrm{\mu m}$. A log of the observations is presented in
Table \ref{obs} and {{Fig.}} \ref{uv_cov} displays the projected
baselines for the observations of HD 327083. While the use of FINITO
can bias the final visibilities, the observations were performed
before the FINITO tracking information was recorded and thus we cannot
assess this possibility.

\smallskip

Data reduction was performed with the amdlib software \citep[version
3, see][]{Tat-amdlib,Chelli2009}. The standard data reduction
procedure, i.e. determining the pixel-to-visibility-matrix and
converting the observed fringe patterns to measurements of the
coherent flux, was followed. The data were then subject to frame
selection in which the frames were ranked based on their fringe
signal-to-noise ratio (SNR) and the best 5 per cent were selected. We
found that retaining more than 5 per cent of the frames resulted in
inconsistent visibilities for two baselines with similar lengths and
position angles (UT1-UT2 on the $\mathrm{13^{th}}$ and UT2-UT3 on the
$\mathrm{14^{th}}$). Therefore, we adopted a selection criterion of
the best 5 per cent of frames. The difference between visibilities
generated with a selection criterion of 5 and 20 percent was generally
less than 0.05 for other baselines, and was typically comparable to
the uncertainties. 

\smallskip

The small selection criterion resulted in relatively noisy phase
information. Therefore, differential phases were calculated with a 50
per cent selection criterion. Even with this enhanced selection rate,
the closure phase measurements observed with the UT1-UT2-UT3
configuration exhibited large uncertainties. To reduce the
uncertainties, these measurements were recalculated with a selection
rate of 80 per cent. The observations with the UT2-UT3-UT4
configuration reveal a non-zero closure phase. This exhibits a slight
trend with the frame selection rate, decreasing at high frame
selection rates. To minimise this effect, we recalculated these
measurements with a frame selection rate of 10 per cent.

\smallskip

Observations of standard stars (typically with diameters of
$\sim$1~mas) were used to create the visibility transfer
function. This was done by dividing the visibilities of the standards
at each spectral bin by the visibility of a uniform disk with a
diameter equal to that of the standards. Standard stars were observed
before and after the observations of HD 327083. This allowed the
behaviour of the transfer function with time to be
constrained. Nonetheless, there could be a systematic uncertainty in
the transfer functions of the order of 0.05 due to temporal
variations. The transfer functions were generally of the order of 0.1
indicating poor fringe contrast, perhaps as a result of telescope
vibrations. 

\smallskip

The standard stars used are late type giants with spectral types of
K3III, G8III/IV and K4/5 III\footnote{from SIMBAD:
http://simbad.u-strasbg.fr/simbad/}. Such stars exhibit photospheric
absorption features due to molecules. For example, both G and K type
giants exhibit absorption due to CO at $\mathrm{\sim2.3 \mu m}$. The
visibilities of the standard star HD 161068 change over the
wavelengths corresponding to the CO absorption. This implies that we
begin to resolve the outer layers of the stellar surface. To avoid
introducing artifacts in the science data, the transfer functions were
fit with low order polynomial functions which were then used to
calibrate the science data.

\smallskip

The spectrum of HD 161068 was then used to remove telluric
absorption features from the spectrum of HD 327083. To achieve this
without {{introducing}} artifacts in the final spectra, the intrinsic CO
bandhead absorption features in the spectrum of HD 161068 also had to
be removed. This was done following the example of
\citet{Tatulli2008}. A template spectrum was created by averaging
spectra of stars with a similar spectral type in the database
of \citet[][HD 62721, HD 70272 and HD 164058]{Wallace1997}. The
average spectrum was then multiplied by the slope of a model SED
\citep[taken from][]{Cast_mod}, which was computed for parameters
appropriate for the spectral type of the standard. The resultant
spectrum was smoothed to match the resolution of the observations and
was then used to remove the intrinsic absorption features of the
standard spectra.

\begin{center}
  \begin{table}
    \begin{center}
      \caption{Observations of HD 327083.\label{obs}}
      \begin{tabular}{l l  r r r}
        \hline
        Date & \multicolumn{2}{c}{Baseline} & \multicolumn{1}{c}{PA} &  \multicolumn{1}{c}{SNR} \\
        &    (UTs)        &\multicolumn{1}{c}{(m)}       &\multicolumn{1}{c}{($\mathrm{\degr}$)} & \multicolumn{1}{c}{(mean)}\\
        \hline
        \hline
        13/04/09 & UT1-UT2 & 52.3 & 36.1 & 1.9\\
        13/04/09 & UT2-UT3 & 42.8 & 51.3 & 4.2\\
        13/04/09 & UT3-UT1 & 94.2 & 42.9 & 0.9\\
        14/04/09 & UT2-UT3 & 46.5 & 32.0 & 6.1 \\
        14/04/09 & UT3-UT4 & 59.6 & 100.2 & 4.1 \\
        14/04/09 & UT4-UT2 & 88.1 & 70.8 &   1.6\\
        $\,$\\
        13/04/09 & \multicolumn{4}{l}{{{Standard stars: HD 194013 \& HD 161068}}}\\ 
        14/04/09 & \multicolumn{4}{l}{{{Standard stars: HD 109963 \& HD 161068}}}\\ 
        \hline

      \end{tabular}
      \tablefoot{{Column 1 contains the dates of the observations,
          columns 2, 3 and 4 describe the baselines used and column 5
          lists the average frame signal to noise ratio (SNR) for each
          baseline. The adopted diameters of the calibrators were HD
          194013: 1.1132$\pm$0.091, HD 161068: 1.43972$\pm$0.01975, HD
          109963: 0.784989$\pm$0.01025~mas.}}
    \end{center}
  \end{table}
\end{center}

\section{Results}

\label{res}

\begin{center}
  \begin{figure}
    \begin{center}
      \includegraphics[width=0.5\textwidth]{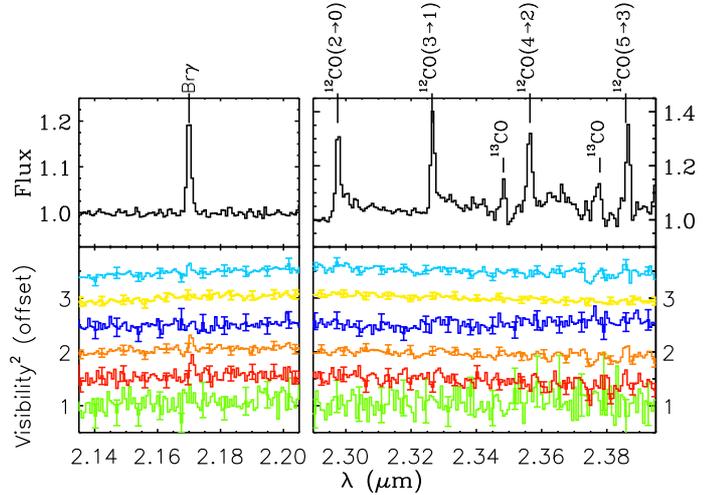}
      \caption{The average spectrum and differential visibilities of
        HD 327083. The upper panel presents the spectrum featuring
        Br$\gamma$ and CO emission. The visibilities are shown offset
        (by an integer times 0.5) to aid comparison. They are shown to
        scale in {{Fig.}}  \ref{vis_sq}.\label{spectrum}}
    \end{center}
  \end{figure}
\end{center}

\begin{center}
  \begin{figure}
    \begin{center}
      \begin{tabular}{l }
        \includegraphics[width=0.4\textwidth]{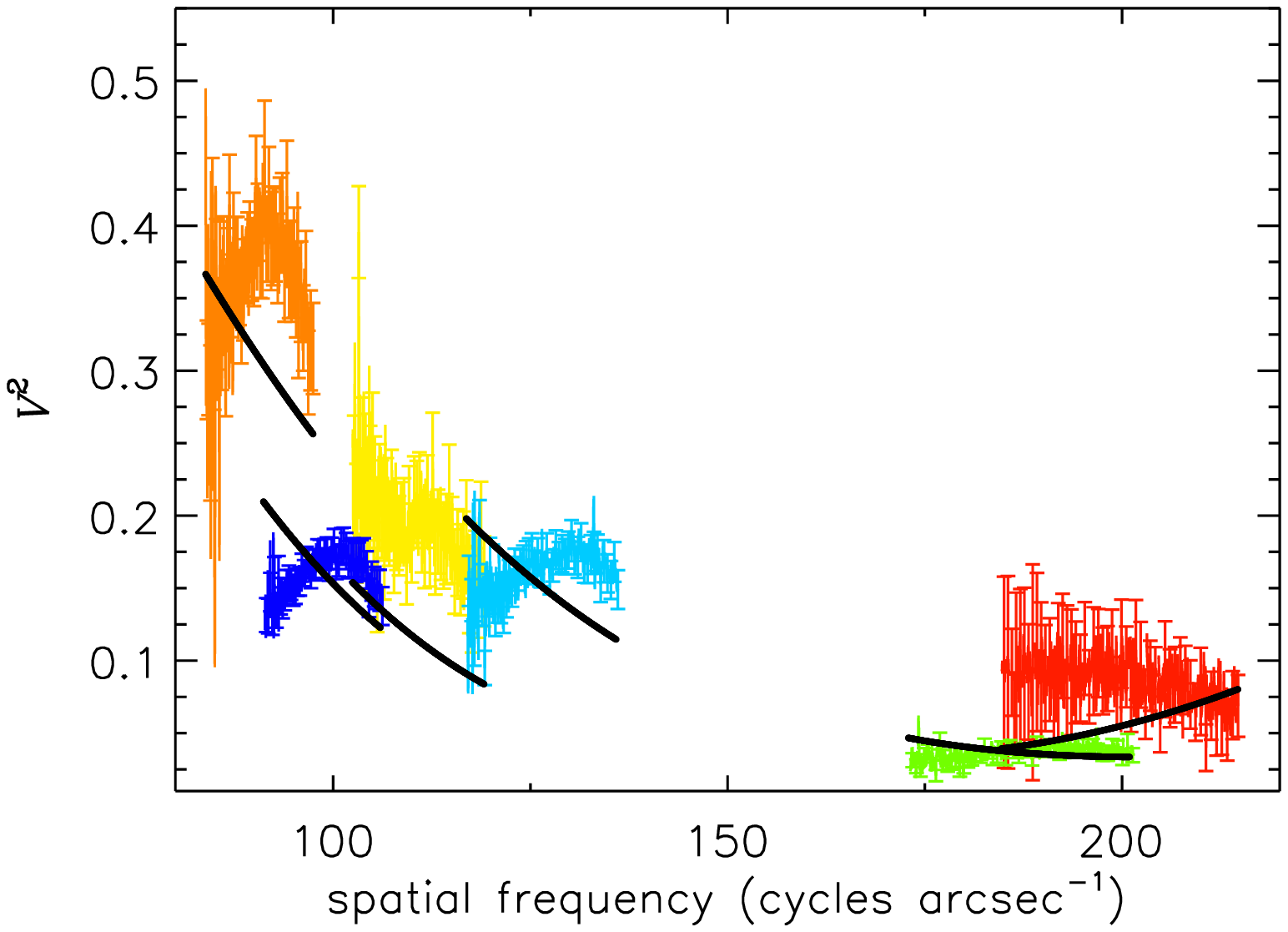}  \\
        \includegraphics[width=0.4\textwidth]{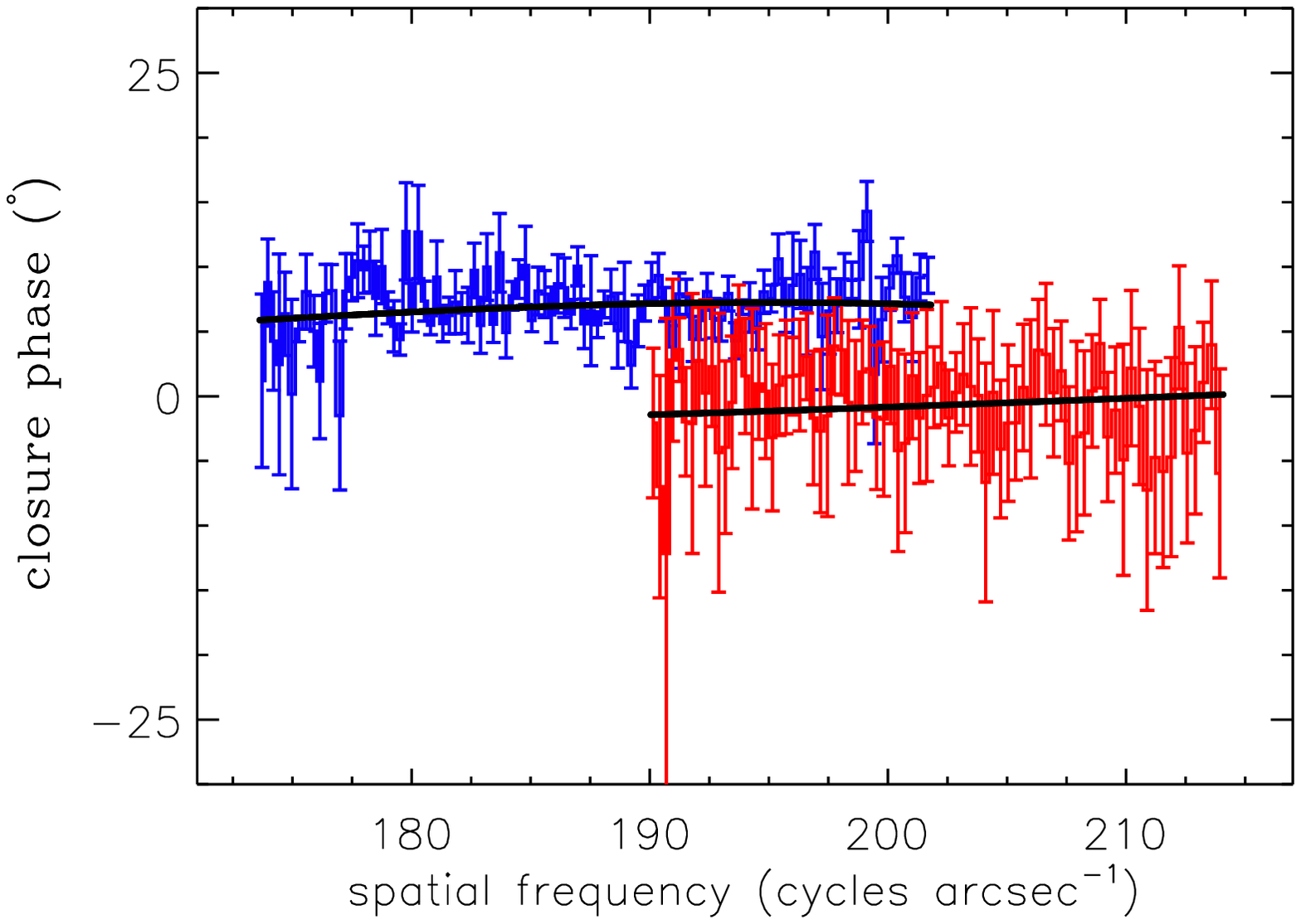}  \\
      \end{tabular}
      \caption{{The visibilities and closure phases of HD 327083 as a
          function of spatial frequency. The closure phase
          measurements have been re-binned by a factor of 5. Closure
          phases observed with the UT1-UT2-UT3 configuration at
          spatial frequencies below 190 cycles per arcsecond have been
          omitted as these exhibited a large scatter. The solid lines
          represent the best fitting model discussed in {{Sect.}}
          \ref{bin_mod}. \label{vis_sq}}}
    \end{center}
  \end{figure}
\end{center}

The observations of HD 327083 are presented in Figs \ref{spectrum} and
\ref{vis_sq}. The spectrum, displayed in {{Fig.}} \ref{spectrum},
exhibits Br$\gamma$ emission and CO overtone bandhead emission at
$\mathrm{\sim 2.3 \mu m}$. While the $\mathrm{^{12}}$CO overtones
dominate the emission, the spectrum also exhibits two features at
$\mathrm{\sim2.35}$ and $\mathrm{\sim2.37~\mu m}$ that are identified
as $\mathrm{^{13}CO}$ emission {{\citep[these can also be seen in the
    lower resolution data of][]{McGregor1988}}}. This is consistent
with the identification of HD 327083 as a star in an advanced
evolutionary state and confirms the object's classification as a
sgB[e] star \citep[see][]{Kraus2009CO,Liermann2010}. The Br$\gamma$
emission has a Full-Width-at-Half-Maximum of
$\sim$220~$\mathrm{km\,s^{-1}}$, indicating that it is marginally
resolved (in the spectral domain).

\smallskip

The visibilities of HD 327083 as a function of wavelength and spatial
frequency are presented in Figs \ref{spectrum} and \ref{vis_sq}
respectively. The squared visibilities range from $\sim$0.05 to
$\sim$0.4. Consequently, we find that the source of {{the NIR continuum
is resolved on all baselines}}. We note that a slight increase in
visibilities is detected across the Br$\gamma$ emission at certain
position angles ($\mathrm50-100^{\circ}$, see {{Fig.}}
\ref{spectrum}). No clear change is observed in the visibilities
across the CO bandhead emission. There is a suggestion of an increase
in visibilities over the $\mathrm{4^{th}}$ bandhead, but as this is
not replicated over the adjacent bandhead, we do not treat this as a
robust detection. The increase in visibilities indicates that in
certain directions, the Br$\gamma$ emitting region is less extended
than the region responsible for the continuum emission. The Br$\gamma$
and CO emission features are of a comparable strength, so this
difference in behaviour is unlikely to be due to a difference in
contrast with the continuum. Therefore, the visibilities also indicate
that the spatial distribution of the CO and the Br$\gamma$ emission is
different.

\smallskip

\begin{center}
  \begin{figure}
    \begin{center}
      \begin{tabular}{l }
        \includegraphics[width=0.325\textwidth]{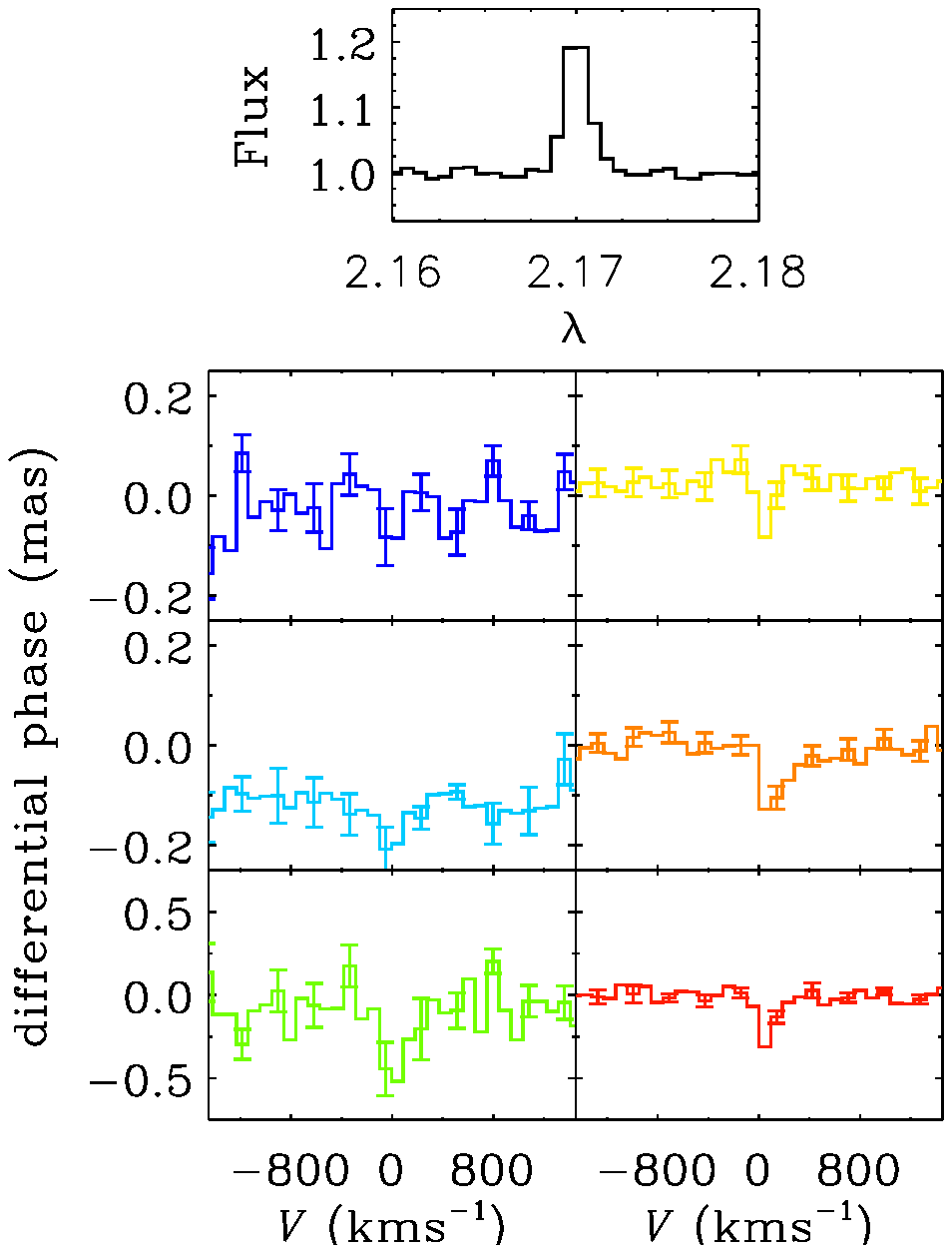} \\
        \includegraphics[width=0.325\textwidth]{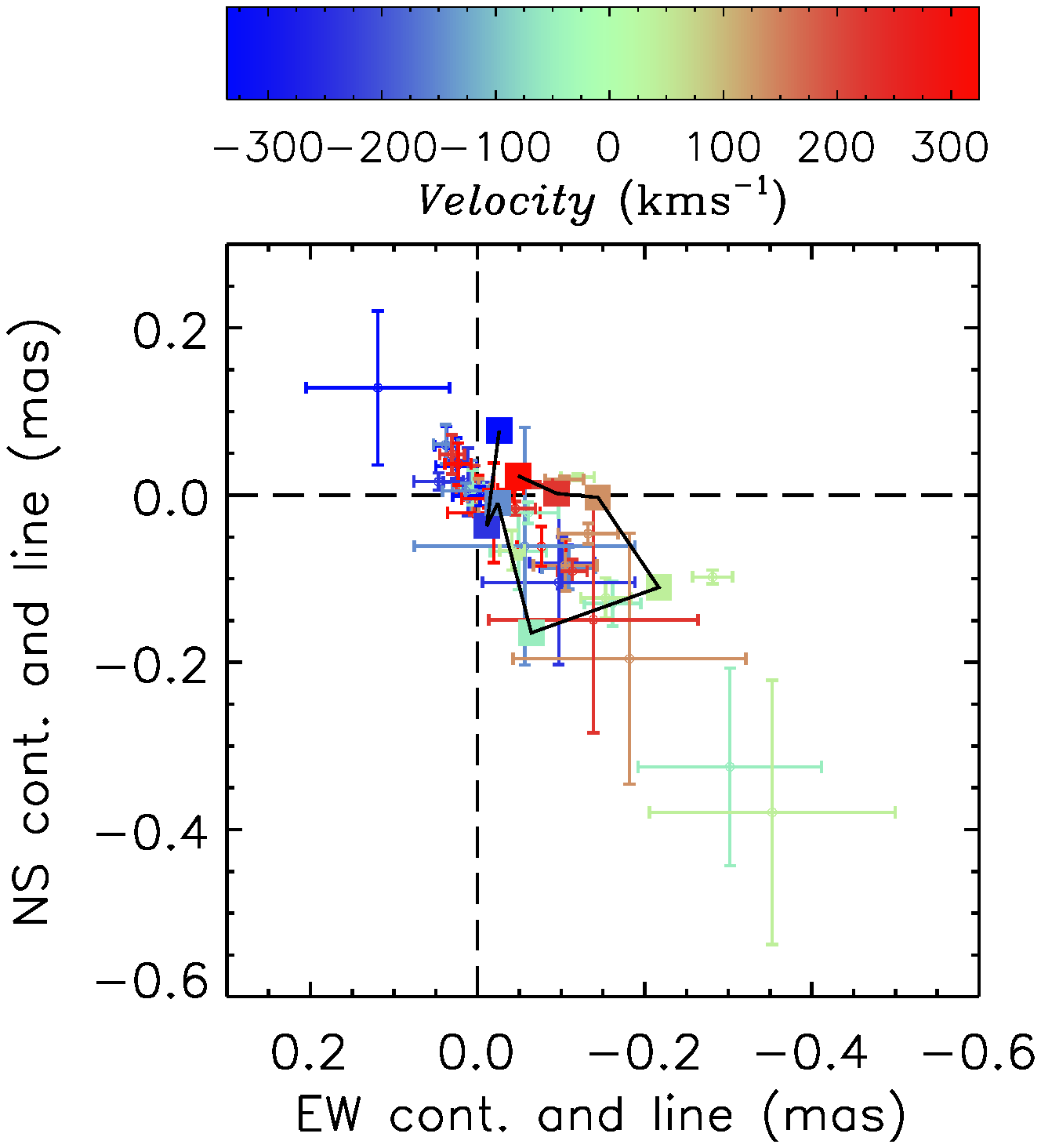} \\
      \end{tabular}
      \caption{The average differential phases over the Br$\gamma$
        emission for each of the six baselines used and the calculated
        photo-centre shift. The calculated total shift per spectral
        channel is marked by the square points.\label{dp}}
    \end{center}
  \end{figure}
\end{center}

{{The average differential phase measurements over the Br$\gamma$
emission are shown in {{Fig.}} \ref{dp}. Changes in phase are
associated with a displacement of the photocentre via the equation: 
\begin{equation} p= -\frac{\Delta \phi}{2\pi}\cdot
  \frac{\lambda}{B} \end{equation}\ where $\lambda$ and $B$ are the
wavelength and projected baseline associated with the observation, $p$
is the projected photocentre displacement and $\Delta \phi$ is the
differential phase signature \citep[see e.g.][]{Lachaume2003}. The
sign of the phase shift determines the direction of the photocentre
displacement. We re-reduced the AMBER data of \citet{Benisty_ZCMa},
who map the differential phase signature of Z CMa onto the direction
of a known jet, and find our treatment of the differential phases is
consistent with theirs. The final photocentre shift per spectral
channel is calculated by forming a system of linear equations from the
linear shifts in the directions of the individual baselines.}}

\smallskip

A shift in phase corresponding to a photo-centre shift of the order of
0.1~mas can be observed over the Br$\gamma$ emission, particularly in
the data obtained with the UT2-UT3-UT4 configuration. The one-sided
nature of the differential phase signature indicates that the
direction of the photo-centre shift is constant. This is shown in
{{Fig.}} \ref{dp} where the displacements are shown on the plane of
the sky. The excursion in one direction is contrary to the case of a
rotating disk where the blue and red shifted photo-centres are located
on opposite sides of the continuum. No such signature is observed over
the CO bandhead emission. This indicates that it has a more
symmetrical and/or more compact distribution.

\smallskip

Turning to the closure phase measurements, also shown in
Fig. \ref{vis_sq}, the averaged measurement for each configuration
reveals a closure phase that is essentially constant with wavelength
(spatial frequency). In the case of the UT1-UT2-UT3 configuration, the
average closure phase is $\sim$0$\mathrm{^{\circ}}$ while in the case
of the UT2-UT3-UT4 configuration it is approximately
7$\mathrm{^{\circ}}$. The closure phases of the calibrators do not
exhibit this signature and thus it is unlikely to be due to a
systematic effect. Instead, the non-zero closure phase observed
suggest the presence of an asymmetric flux distribution, which is
expected in the case of a binary system.

\subsection{Modelling the observations}

\subsubsection{Continuum visibilities and closure phases}

\label{bin_mod}

To investigate the circumstellar environment traced by the
observations, we fit the visibilities and closure phase measurements
with a simple geometrical model.

\smallskip

The model is constructed to represent the binary system detected by
\citet{Miro2003}. Specifically, the binary components are treated as
uniform disks with radii of 27 and 86~{{$R_{\odot}$}}, diameters of
0.17 and 0.53~mas at 1500~pc, respectively. A flux ratio of 3.5 was
estimated from the visual brightness ratio and model SEDs \citep[taken
from][]{Cast_mod} appropriate for the stellar properties reported by
\citet{Miro2003}.

\smallskip

\begin{center}
  \begin{figure}
    \begin{center}
      \includegraphics[width=0.4\textwidth]{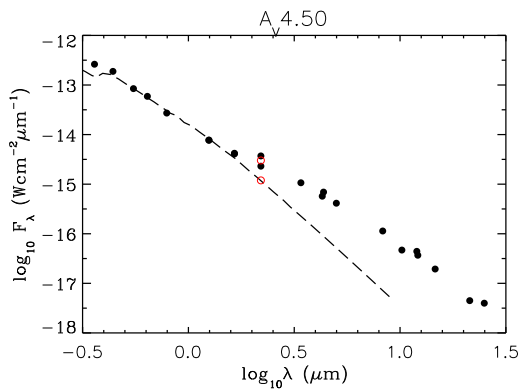} 
      \caption{The de-reddened SED of HD 327083. The dashed line marks
        the model SED of the proposed binary system, scaled to fit the
        visible SED. The model SEDs are taken from
        \citet{Cast_mod}.\label{sed_fig}}
    \end{center}
  \end{figure}
\end{center}

The observed visibilities are unlikely to be fit by a model of a
binary system alone. Such a scenario would result in a sinusoidal
visibility signature, and the low visibilities observed clearly
indicate an extended component. The system also contains circumstellar
material, as evidenced by the infrared excess. Therefore, we add an
additional component to the binary model.  Determining the
circumstellar contribution to the $K-$band flux requires the SED of HD
327083. Using the data presented in: \citet{Miro2003},
\citet{Cutri2003}, \citet{MSX} and \citet{IRASPS}, the SED of HD
327083 was constructed (see {{Fig.}} \ref{sed_fig}).  We adopted a value
of $A_V$ = 4.5 to de-redden the final SED. Based on optical colour
indices, \citet{Miro2003} estimate a value of $A_V = 5.6
\pm0.5$. However, the value adopted provides a closer match between
the UV flux and the model binary SED. The final estimate for the
circumstellar contribution to the $K-$band flux is 1.5 times that of
the binary. If the {{reddening}} is greater than the value adopted, this is
reduced and vice versa.

We initially represented the material by an elliptical
Gaussian function centred on the position of the primary
component. Fitting the visibilities only, the best fitting model
resulted in $\chi^2=4.4$ and featured a Gaussian that was extended
along a PA of $\sim$160$\mathrm{^{\circ}}$ with a major FWHM of
approximately 10~mas. As a result, it is significantly more extended
than the expected binary separation, which is limited to a maximum of
several mas by radial velocity variations. Therefore, it would appear
{{that}} the circumstellar material is not truncated by the presence
of the companion. This presents two possibilities: either the NIR
excess traces a bi-polar outflow from the primary that is {{oriented}}
perpendicular to the binary system or the NIR excess originates in a
circum-binary disk.

\smallskip

The zero-closure phase observed with the UT1-UT2-UT3 array, which is
an approximately linear configuration with a PA of
$\mathrm{40^{\circ}}$, implies that the binary PA is
$\mathrm{\sim130^{\circ}}$. As a result, taking into account the
maximum expected binary separation, it appears the binary system is
situated interior to the source of the NIR excess. This is more
reminiscent of a circum-binary disk than an outflow. Therefore, we now
model the circumstellar material as a circum-binary disk, which we
approximate as an elongated, uniform ring. We note that the
circumstellar material traced by the line emission and the NIR
continuum is not necessarily distributed identically. Therefore, to
begin with, we consider only the continuum light and fit the
visibilities over the entire spectral range. The visibilities over the
Br$\gamma$ and CO emission are discussed in the next section.

\smallskip

The free parameters of the final model were: the binary separation and
position angle (PA), the radius and {{width}} of the uniform ring, its
elongation and its PA. To fit the observed visibilities with the
model, we used LITpro\footnote{LITpro software available at
  http://www.jmmc.fr/litpro} ({{Lyon}} Interferometric Tool
prototype), a piece of software developed by the Jean-Marie Mariotti
Center {{\citep[JMMC,][]{LITpro}}}. LITpro fits geometrical models
to interferometric observables by using the Levenberg-Marquardt
algorithm to minimise $\chi^2$. Since the radial velocity variations
of HD 327083 suggest the maximum binary separation is of the order of
several $AU$, the separation of the binary model was limited to less
than approximately 4.0~mas.

\smallskip

\begin{center}
  \begin{figure}
    \begin{center}
      \includegraphics[width=0.35\textwidth]{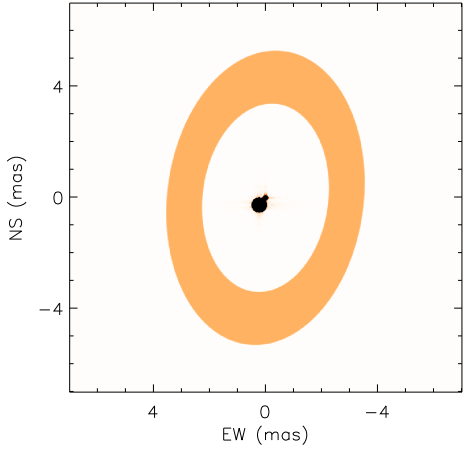}
      \caption{The image of the best fitting model. North is towards
        the top of the page and East is to the left. The image has
        been logarithmically scaled.\label{image}}
    \end{center}
  \end{figure}
\end{center}

In an attempt to identify the global minima in $\chi^2$, we
initialised the fitting program with 50 different sets of values for
the free parameters. The best fitting model resulted in a $\chi^2$
value of 4.4. As can be seen in {{Fig.}} \ref{vis_sq}, this moderately
high value is partly due to the inability of the binary model to
reproduce the curve of the visibilities at low spatial frequencies
($\sim$100 cycles per arcsec). The close binary system cannot produce
modulations in the visibilities with such a pronounced curve. This may
be the signature of some additional source of flux such as localised
nebulosity or an additional companion that is fully resolved at
baselines of $\sim$90m. Further observations are planned to improve
our $u,v$ coverage to investigate this. For now, we surmise that the
value of $\chi^2$ = 4.4 indicates that the fit is reasonable for the
model that we are using. The best fitting parameters are presented in
Table \ref{bf_pars} and the visibilities of the best fitting model are
displayed in {{Fig.}} \ref{vis_sq}. Finally, the image of the best
fitting model is show in {{Fig.}} \ref{image}.

\begin{center}
  \begin{table}
    \caption{The best fitting parameters of the binary and elongated ring model from LITpro.\label{bf_pars}}
    \begin{center}
      \begin{tabular}{l l |  l }
        \hline
        Parameter & Value & Notes \\
        \hline

        Bin. sep &  $\mathrm{0.34\pm0.01}$~mas & 0.5~$AU$, 4.0~$R_{\star}$\\

        Bin. PA &   $\mathrm{139.1\pm0.1}$$\degr$\\

        Ring radius&    $\mathrm{3.40\pm0.02}$~mas & 5.1~$AU$, 40.5~$R_{\star}$\\

        Ring width &   $\mathrm{1.91\pm0.03}$~mas & 2.9~$AU$, 22.7~$R_{\star}$ \\

        Ring flattening &   $\mathrm{1.51\pm0.01}$ &  inclination $\sim$48.5 \\

        Ring PA &  $\mathrm{173.0\pm0.2}$$\degr$\\

        $\chi^2$ & 4.44  \\

        \hline
      \end{tabular}
      \tablefoot{{As discussed in \citet{DS2011}, the uncertainties
          reported by the Levenberg-Marquardt algorithm should be
          considered lower limits to the actual uncertainty in the
          best fitting parameters. The sizes in $AU$ and $R_{\star}$
          were calculated assuming a distance of 1500~pc.}}
    \end{center}
  \end{table}
\end{center}

\subsubsection{Br$\gamma$ and CO visibilities}

At certain position angles, the visibilities of HD 327083 increase
over wavelengths corresponding to the Br$\gamma$ emission (see {{Fig.}}
\ref{spectrum}). The increase in visibilities is accompanied by a
differential phase shift indicating an offset from the continuum
photo-centre. The CO bandhead emission, which has a higher
line-to-continuum ratio, exhibits no such features. The fact that the
CO emission is not associated with such signatures strongly suggests
that they are not an artifact caused by the frame selection
process. Here we investigate what geometry can recreate this
behaviour.

\smallskip

The visibility over a spectral line can be calculated by:\\
\begin{equation}V_{\rm{line}} = \frac{F_{\rm{tot.}}V_{\rm{tot}} -
    F_{\rm{cont.}}V_{\rm{cont.}}}{F_{\rm{line}}}\label{eq_1} \end{equation}
where $F_{\rm{line}}$=$F_{\rm{tot.}}-F_{\rm{cont.}}$ and assuming that the phase is constant over the line \citep[see e.g.][]{Malbet2007}. If there is a change in the differential phase of $\Phi^\prime$, the following equation must be used:\\
$V_{\rm{line}} =$
\begin{equation}\frac{\sqrt{|V_{\rm{tot.}}F_{\rm{tot.}}|^2 + |V_{\rm{cont.}}F_{\rm{cont.}}|^2 - 2V_{\rm{tot.}}F_{\rm{tot.}}V_{\rm{cont.}}F_{\rm{cont.}}\cos\Phi ^\prime}}{F_{\rm{line}}}\label{eq_2}\end{equation}
\citep[see e.g.][]{Weigelt2007}. There is a clear differential signature over the Br$\gamma$ emission. Therefore, to calculate the visibility over this line we employ equation \ref{eq_2}. The CO bandhead emission exhibits no such signature and thus we use equation \ref{eq_1} to calculate the CO visibilities. Since the lines are only marginally resolved in the spectral domain, we calculate the visibilities for the line peaks only. In the case of the CO emission, we concentrate on the $\rm{1^{st}}$ bandhead alone. We list the final quantities in Table \ref{line_vis}. The ratios of the line and continuum visibilities indicate that the line emitting regions are generally more compact that the continuum emitting region.

\begin{center}
  \begin{table}
    \begin{center}
      \caption{The visibilities over the spectral lines and the ratio
        of the line visibilities with those of the adjacent
        continuum.\label{line_vis}}
      \begin{tabular}{l c c c c c c}
        \hline
        Region &  \multicolumn{6}{c}{$V^2_{\rm{final}}$ }\\
        & UT1-2 & UT2-3& UT1-3 & UT2-3 & UT3-4 & UT2-4\\
        \hline

        \hline
        Br$\gamma$ & 0.20&0.70&0.27&0.19&0.37&0.14\\
        Br$\gamma$/cont & 1.22&1.93&3.60&1.13&2.12&3.64 \\

        CO & 0.23&0.49&0.07&0.19&0.19&0.03\\
        CO/cont & 1.21&1.25&0.82&1.13&1.16&0.79 \\
        \hline
      \end{tabular}
      \tablefoot{{The visibilities correspond to the 6 average baselines presented in {{Fig.}} \ref{uv_cov}.}}
    \end{center}
  \end{table}
\end{center}

To estimate characteristic sizes of the regions responsible for the
observed emission lines, we attempt to recreate the line visibilities
presented in Table \ref{line_vis} with a simple model of a flattened
ring. To assess the relationship between the line and continuum
emission, we attempt to fit the line visibilities with the same
elongation and PA as the best fitting continuum model and we vary only
the size to match the line visibilities.

\smallskip

The increase in the visibilities over the Br$\gamma$ line at PAs of
$\mathrm{50-100^{\circ}}$ clearly indicates that the line emitting
region is more compact than the continuum emitting region. In
addition, it also appears more compact than the CO emission at these
position angles. This decrease in size at specific PAs cannot be
recreated with the simple ring model with the same flattening and PA
as the continuum ring. On the contrary, the CO visibilities can be
relatively well reproduced with a family of ring models which have an
inner radius in the range $\sim$0.5--2.5~mas and a total extension
(inner radius plus width) of approximately 3.5~mas (5.25~$AU$). This
corresponds to the inner radius of the ring that fits the continuum
visibilities and thus places the CO emission within the continuum
emitting region. However, we note that without a clear visibility
signature, the size of the CO emission is not tightly
constrained. Therefore, while we find that a compact CO emitting
region is consistent with the data, a more extended region cannot be
excluded.

\subsection{A summary of the observational results}

To summarise the results, we have shown that the observations can be
reproduced by a simple model of a close binary system with an
additional circum-binary component. We have resolved the material
responsible for the NIR excess of HD 327083 and its line emission. The
continuum and CO emission appear to originate from an elongated
disk-like structure that encompasses the binary system. The Br$\gamma$
emission is distributed differently and appears more compact than the
continuum and CO emission. These results constitute the first direct
constraints on the distribution of the circumstellar material
associated with HD 327083 and allow a new insight into the behaviour
of this object and its classification as a sgB[e] star. This is
discussed in the following section.

\section{Discussion}

\label{disc}

\citet{Miro2003} report that HD 327083 is a spectroscopic binary with
a period of approximately 6~months and an orbital semi-major axis of
$\sim$2~$AU$. Our observations can be reproduced by a model of a
binary system. In particular, the non-zero closure phase detected is a
clear indicator of an asymmetric flux distribution which we attribute
to the presence of the binary companion. This will be investigated
further with multi-epoch observations.

\smallskip

It has been suggested that the system experiences mass transfer during
phases of the orbit close to periastron \citep{Miro2003}. If this is
the case, the circumstellar material may not be mass lost from the
primary via an equatorial outflow (as envisaged in the standard sgB[e]
scenario). Instead, it is possible that the material is lost from one
of the components as it fills its Roche lobe. Our high spatial
resolution observations resolve the circumstellar material surrounding
the system and constrain the binary separation. This allows us to
examine the relationship between them in order to assess the
suggestion that the circumstellar material is the result of binary
interactions.

\smallskip

\smallskip

The continuum visibilities and closure phases suggest that the binary
is surrounded by a dusty disk. While the extent of the CO bandhead
emission is not well constrained, the data are consistent with the CO
emission being aligned with the continuum emission although less
extended. CO bandhead emission originates in hot, dense gas, exactly
the conditions expected in circumstellar disks \citep[][and references
therein]{meCO}. Therefore, the inferred distribution of the CO
emission is consistent with the scenario of a gaseous disk interior to
a larger, dusty disk. The inner radius of the continuum emitting ring,
$\sim$40~$R_{\star}$, is compatible with the expected dust sublimation
radius of an early B-type star, which supports the scenario of an
outer dusty disk. Turning to the Br$\gamma$ emission, an increase in
visibilities at certain PAs suggest a more compact and more asymmetric
distribution than those of the continuum and the CO emission. A
plausible scenario to account for this is that the Br$\gamma$ emission
contains a contribution from the circumstellar environment close to
the central star, where CO molecules cannot exist.

\smallskip

The differential phase signature associated with the Br$\gamma$ line
and the lack of a signature over the CO bandhead emission is
consistent with the scenario in which the CO and Br$\gamma$ emission
originate in different locations. The lack of a differential phase
signature over the CO emission is attributed to the small photo-centre
offsets of CO emission originating in disks \citep[see e.g.][]{meCO}
and the limited spectral resolution of AMBER in the medium resolution
mode. High spectral resolution observations are planned to assess this
hypothesis. The one-sided nature of the photo-centre offset over
Br$\gamma$ indicates that it originates from an asymmetric
region. While a scenario in which the Br$\gamma$ emission emanates
directly from one of the stellar components of a binary system could
result in such a signature, the PA of the photocentre shift appears to
differ from the inferred binary PA. Therefore, it seems that the
Br$\gamma$ emission traces some additional asymmetry in the
environment of HD 327083. We note that the presence of the close
binary companion might result in an asymmetric environment on
mas scales which could explain the appearance of the differential
phase signature.

\smallskip

To summarise, the interferometric data-set are consistent with the
scenario of a circum-binary disk. Since HD 327083 is an evolved
system, the circum-binary material is presumably composed of mass lost
from the system. In the case of sgB[e] stars, it is assumed that the
circumstellar material is the result of intrinsic mass loss. However,
here the mass loss could also be the result of binary interactions. If
this is the case, binary sgB[e] stars may represent a distinct
sub-sample of the sgB[e] population \citep[similar to lower luminosity
binary Be stars, see e.g.][]{Miro2007}. Here, we assess this notion by
comparing these observations to similar studies of B[e] and A[e]
objects.

\smallskip

Few similar studies exist. However, we note that
CPD$-$57$\mathrm{^{\circ}}$2874, one of the few other sgB[e] stars
studied with AMBER, is not thought to be a binary and does exhibit a
different behaviour to HD 327083. In the data presented here, the
visibilities increase over the Br$\gamma$ line, indicating that the
line emitting region is smaller than that emitting in the
continuum. In the case of CPD$-$57$\mathrm{^{\circ}}$2874, the reverse
is true \citep[see][]{deSouza2007}. We also note that the scenario of
a dusty circum-binary disk with a gaseous interior is similar to the
circumstellar environment of the A[e] star HD 62623 as revealed by
VLTI/AMBER observations \citep{Millour2011}. As an A-type star, HD
62623 is cooler than sgB[e] stars and is also less luminous. The
currently favoured explanation of the sgB[e] phenomenon, the
bi-stability mechanism, is more effective at high luminosities
\citep{Pelupessy2000}. Consequently, it is thought that the
circumstellar material surrounding HD 62623 is instead the result of
binary interactions \citep[][]{Millour2011}. The similarity between
the circumstellar environments of HD 62623 and HD 327083 supports the
hypothesis that the B[e] behaviour of HD 327083 is the result of
binary interactions rather than intrinsic mass loss.

\section{Conclusion}

\label{conc}

This paper presents VLTI and AMBER observations of the sgB[e] star HD
327083. Here we list the salient findings.

\begin{itemize}

\item[--]{HD 327083 exhibits $\mathrm{^{13}CO}$ bandhead emission,
    this confirms that this object is in a {{post-main}} sequence
    evolutionary phase.}

\item[--]{We spatially resolve the circumstellar environment of HD
    327083 for the first time. A non-zero closure phase is
    observed. We attribute this to the binary companion detected via
    radial velocity variations.}

\item[--]{We find that the spatial distribution of the NIR continuum
    excess emission can be described as an elongated ring 10.6~mas in
    size that encompasses the binary system. We associate this with a
    dusty circum-binary disk.}

\item[--]{The data are consistent with the notion that the B[e]
    behaviour of HD 327083 is due to binarity.}

\end{itemize}

As a final remark, we note that currently few high spatial resolution
observations of sgB[e] candidates have been reported. To date, AMBER
observations of three Galactic sgB[e] candidates have been published:
HD 87643 \citep{Millour2009}, CPD$-$57$\mathrm{^{\circ}}$2874
\citep{deSouza2007} and V921 Sco
\citep{Kraus2008lines,Kreplin2011}. One of these stars is found to
have a binary companion (HD 87643). With these results, the fraction
of sgB[e] candidates observed with high spatial resolution that could
be interacting binaries increases to 50 per cent. This raises the
possibility that binarity may play a more important role in the
Galactic sgB[e] phenomenon than previously thought. High angular
resolution observations of additional sgB[e] stars are required to
assess this possibility.

\begin{acknowledgements}
  The authors acknowledge the helpful and insightful comments of an
  anonymous referee whose suggestions helped improve the paper. This
  research has made use of the \texttt{AMBER data reduction package}
  of the Jean-Marie Mariotti Center\footnote{Available at
    http://www.jmmc.fr/amberdrs}. This publication also makes use of
  data products from the Two Micron All Sky Survey, which is a joint
  project of the University of Massachusetts and the Infrared
  Processing and Analysis Center/California Institute of Technology,
  funded by the National Aeronautics and Space Administration and the
  National Science Foundation. Finally, this research has employed the
  Jean-Marie Mariotti Center \texttt{LITpro} service co-developed by
  CRAL, LAOG and FIZEAU.

\end{acknowledgements}

\bibliographystyle{aa} 
\bibliography{bib}

\end{document}